# Physics Faculty and Educational Researchers: Divergent Expectations as Barriers to the Diffusion of Innovations

## Abstract

Physics Education Research (PER) practitioners have engaged in substantial curriculum development and dissemination work in recent years.  Yet, it appears that this work has had minimal influence on the fundamental teaching practices of typical physics faculty.  To better understand this situation interviews were conducted with 5 likely users of physics education research.  All reported making changes in their instructional practices and all were influenced, to some extent, by educational research.  Yet, none made full use of educational research and most had complaints about their interactions with educational researchers.  In this paper we examine how these instructors used educational research in making instructional decisions and identify divergent expectations about how researchers and faculty can work together to improve student learning.  Although different instructors emphasized different aspects of this discrepancy between expectations, we believe that they are all related to a single underlying issue: the typical dissemination model is to disseminate curricular innovations and have faculty adopt them with minimal changes while faculty expect researchers to work with them to incorporate research-based knowledge and materials into their unique instructional situations.  Implications and recommendations are discussed.

# I. Introduction

In recent decades, Physics Education Research (PER) has developed knowledge about issues related to the teaching and learning of physics as well as successful instructional strategies and materials based on this knowledge. It is unclear, however, what effect these substantial efforts have had on the actual teaching of introductory college-level physics. Evidence from empirical studies[1,2,3] as well as the opinions of prominent National committees (for example, see Refs 7 and 8) and PER practitioners[9-11] all suggest that most physics instructors continue to use traditional teaching practices[12] and that dissemination of reforms is an important unsolved problem. For example, as the rationale for its 2003 report, the Committee on Undergraduate Science Education[8] points to the strong STEM research base on effective teaching approaches and then questions "why introductory science courses in many colleges and universities still rely primarily on lectures and recipe-based laboratory sessions where students memorize facts and concepts, but have little opportunity for reflection, discussion, or testing of ideas?" (p. 1)

To better understand this dissemination problem, we conducted interviews with a purposeful sample of five physics faculty who we believe represent highly likely users of educational research. All reported making changes in their instructional practices and all were influenced, to some extent, by educational research. Yet, none made full use of educational research and most had complaints about their interactions with educational researchers. In this paper we examine how these instructors used educational research in making instructional decisions and identify differences in expectations that appear to be barriers to more full use of educational research. We expect that these barriers are not unique to the instructors in this study.

Elsewhere,[14,15] we describe other results from this study. Most notably that all of the faculty interviewed expressed beliefs about teaching and learning that were more compatible with research-based instructional suggestions than were their self-described instructional practices. When asked about this discrepancy, the instructors cited strong situational constraints that made it difficult to teach in a non-traditional manner. Commonalities such as large class sizes, broad content coverage expectations, classroom infrastructure, scheduling constraints, poor student preparation/motivation, and the institutional reward system all appear to favor traditional instruction. The importance of these situational factors and the associated implications for the PER community cannot be ignored and are discussed elsewhere.[11,14] It also became apparent in the interviews, however, that educational researchers and other physics faculty had different expectations about how the two groups should work together to improve student learning. This discrepancy was expressed directly (and often emotionally) by three of the five instructors we interviewed and indirectly by the other two. Although different instructors emphasized different aspects of this discrepancy, we believe that they are all related to a single underlying issue: the typical dissemination model is to disseminate curricular innovations and have faculty adopt them with minimal changes while faculty expect researchers to work with them to incorporate research-based knowledge and materials into their unique instructional situations.

Divergent expectations are not the only barriers to change. However, unlike many other barriers (situational factors, for example) the PER community has significant control over their interactions with non-PER faculty. Thus, we believe that it is quite useful to explore these divergent expectations and implications they might have for potential changes to curriculum development and dissemination.

## Divergent Expectations

Expectations, of course, are important in all human interactions and divergent expectations often result in conflict. For example, in her work on male-female communication, Debora Tannen found that males and females often have different expectations for conversations about troubles.[16,17] Men tend to expect conversations about troubles to arrive at a solution. Women, on the other hand, tend to expect conversations about troubles to develop intimacy. Thus, when a woman shares her troubles with a man the woman often feels that the man isn't listening because he focuses on solving the problem and doesn't share his own problems. The man becomes frustrated because the woman continues to talk about her problems without an apparent interest in a solving them.

Closer to education, divergent student-teacher expectations may result in conflict when teachers attempt to use innovative instructional methods. For example, students and instructors in science classes often abide by a "hidden contract" whereby students are responsible for sitting quietly and asking clarifying questions while teachers are responsible for presenting clear lectures and solving example exercises that are not too different from test questions.[18] A student who expects to sit passively in such a class will likely be frustrated and resist an instructor who expects class to be interactive.[19] Likewise, the instructor is likely to be frustrated by this resistance.[20]

## II. Participants and Data Collection

The purpose of this study was to understand barriers to instructional change. Because current models of the change process are not able to account for the slow rate of instructional change in college level physics, a primary goal of this project was to generate new categories of barriers.[21] Thus, we used exploratory interviews with a small sample of instructors who have qualities that, according to diffusion-based change models,[22-24] are likely to result in successful instructional change.

### A. Participants

Interviews were conducted with five tenured physics faculty from four different institutions (one small liberal arts college, two regional universities, and one major research university). These faculty had no formal connections with the Physics Education Research (PER) community and were purposefully chosen. We targeted faculty we believed should be ideal consumers of research-based reform. They were all senior faculty with a reputation for being particularly thoughtful and reflective teachers in introductory level physics at their institutions. They all valued teaching, had a strong desire to see their students succeed, and, as discussed later, were all inclined to consider making changes in their instructional practices. If, as is commonly stated, the goal of the physics education reform movement is to create a critical mass of instructors using reformed pedagogical approaches, this type of instructor can be expected to form the core of that critical mass. Thus, any difficulties in interacting with these instructors are very important to understand and remedy.

### B. Interview

Each semi-structured exploratory interview lasted over one hour and contained open-ended questions about instructional goals, current and past instructional practices, attempts to change practices, and familiarity with educational research. For example, questions about instructional change included: (a) How has your practice changed over the course of your career? (b) What has caused you to try new things? (c) What have you tried that you have abandoned? Why? (d) What things make it difficult for you to change? (e) What things have supported your efforts to change? Following each of these general questions, the interviewer (either CH or MD) asked probing questions to elicit specific details about the interviewee's experiences with instructional change. Each interview was transcribed for analysis.

## III. Analysis

In analyzing the interview data we were guided by Clement's levels of knowledge.[21] In this hierarchy, the lowest level, Level 1, is primary-level data. In our study these were individual statements made by the instructors during an interview. Level 2 is observed patterns and empirical laws. In our study these were similarities in ideas expressed in the statements made by different instructors. For example, one pattern we noticed was that the instructors often used basic ideas from educational research, but yet changed these ideas significantly during implementation. Level 3 is the researchers' explanatory models. In our study this is the idea of divergent expectations between change agents and instructors as an important barrier to the diffusion of PER innovations. The goal of an explanatory model is to describe a hidden mechanism that explains the observed patterns. This not only adds explanatory power, but also leads to growth of theory. Finally, Level 4 is formal principles and theoretical commitments. These are the result of repeated testing and refinement of explanatory models in a variety of situations. Thus, we cannot make any Level 4 claims from this initial generative study.

As is common with generative studies, our analysis began with an open coding process of constructing categories of statements from the interview transcripts.[25] Patterns were then sought between and among different categories. Throughout this exploratory process both researchers were engaged in the creation, critiquing, and refinement of the emerging categories. Looking for patterns between categories helped to sharpen the category boundaries. The results section will describe three categories of self-described instructor behavior related to instructional change and four categories related to interactions between the instructors and educational researchers.

## IV. Theoretical Basis

Although the purpose of this study was to generate new ideas, it was conducted and interpreted within the framework of our existing ideas. In this section we propose an adoption-invention continuum which describes important characteristics of possible interactions between educational researchers and other faculty. We also make and support the claim that many change agents operate on the adoption side of this continuum.

**A. Adoption-Invention Continuum**

There are two important participants in the instructional change process. One is the instructors who are interested in or being asked to change their instruction. The other, change agents, are curriculum developers or professional development providers who provide information, materials, encouragement, etc. to help the instructors.

There is a body of literature that explores how these two types of participants interact in the change process.[22-24] Models of the change process typically include at least three activities: (1) instructor becomes aware of a problem with current practice, (2) instructor develops knowledge about a new practice that can minimize or solve the problem, and (3) instructor implements the new practice. There are three basic levels of knowledge that can be developed during the second phase:[23] "awareness" knowledge (knowledge that the instructional strategy exists), "how-to" knowledge (basic knowledge about how to use the strategy properly), and "principles" knowledge (knowledge about why the strategy works – essential for solving unexpected problems that occur during use). Although much of the literature on educational change deals with instructional changes that are developed and disseminated by external change agents, it is important to note that innovations do not necessarily come from external sources, but may be developed entirely by an instructor.[26] We believe this observation should be accounted for in theories of change.

We have identified four basic categories of change that vary in terms of the roles of the external change agent and the instructor in the change process (Figure 1). These are not discrete categories, but rather occur on a continuum. We have found it useful, however, to use these category labels to represent general locations along the continuum. Notably, the responsibilities of the change agent and instructor change significantly as one moves across the continuum (Figure 2). At the adoption pole the change agent develops all of the materials and procedures and gives them to the instructor to implement as is. In its extreme, this pole represents a change agent view that the instructor is irrelevant. At the invention pole the instructor develops everything with minimal external influence. In its extreme, this pole represents an instructor view that educational research is irrelevant. Under adaptation and reinvention, the general idea of a new instructional strategy comes from an external source, but the instructor is responsible for developing important aspects of the strategy. Although it is possible for an instructor to develop these aspects of the strategy with the assistance of a change agent, typically the instructor develops these aspects of the strategy on their own. These instructor-developed principles and details are not always consistent with "best practices" as identified in the educational research literature.[26-31]

## B. Change Agents Expect Adoption/Adaptation

In the remainder of this paper we will examine the expectations that our sample of five physics faculty have about their interactions with change agents. However, it is important to first examine the expectations that change agents have. Although, there are certainly a wide variety of change agents with a wide variety of expectations, it appears that most change agents operate near the adoption/adaptation end of the continuum.[32] This is evident in much of the discourse related to educational change that focuses substantial efforts on developing and testing specific instructional innovations. Once proved successful by their developer, these innovations are then disseminated to instructors who are expected to use them with fidelity. The instructor is not an important part of the development of these strategies and, in fact, is often considered to be a barrier to educational change.[8,33-36] As an example, consider the model of curriculum development and dissemination advocated by the NSF-CCLI program (Figure 3). This model shows the change agent responsibilities as conducting research, developing materials and then helping faculty develop expertise in using these materials.[37]

There is nothing inherently wrong with this perspective. A change agent might imagine that the adoption model would be most effective since it places much of the burden on the change agent to develop innovative strategies and materials. This takes considerable time and expertise that typical faculty may not possess. It would be reasonable to assume that faculty expect this sort of interaction with educational researchers. In addition, the adoption model is important for researchers who are attempting to determine the efficacy of a new instructional method. It is difficult to draw conclusions unless all of the participating faculty are doing more or less the same things.

In spite of these reasons from the change agent perspective that support change agent-instructor interactions on the adoption/adaptation end of the continuum, it is important to understand how instructors perceive their actual and desired interactions with change agents.

## V. RESULTS

In this section we present our findings regarding the experiences of the faculty with instructional change followed by a discussion of the perceptions of the faculty regarding the educational research community and products.

### A. Behavior Related to Instructional Change

As discussed earlier, the existing literature on instructional change suggests useful categories for describing instructor behavior related to instructional change. We categorized each instructor's self-described instructional changes in terms of three basic activities, (1) identifying a problem with existing instruction, (2) becoming aware of or developing a potential solution, and (3) implementing the solution. Thus, we can examine the extent to which each activity is accomplished and the interactions between instructor and change agent that facilitate or hinder each activity.

*Identifying problems with practice*

All of the instructors felt that they faced instructional problems that could, at least potentially, be improved via changes in their instructional practices (Table 1). Although these instructional problems appear to be largely consistent with problems identified by educational research, it was seldom clear from the interviews how these instructors first became aware of these problems -- whether they identified the problems on their own or with the help of educational research. In most cases, though, the instructors report that their belief in the importance of and their understanding of the instructional problems has been enhanced through their interactions with educational researchers. For example, Mary describes "always" having the philosophy that students don't get much from a traditional lecture and that class should be more interactive (Mary 70)[38]. Yet, she describes not realizing the gravity of the situation until giving the Force Concept Inventory[39] (FCI): "*Just the fact that somebody can go through the entire class and still think that you needed a force to cause motion was an eye opener.*" (Mary 233-234).

*Becoming Aware of Research-Based Instructional Methods*

Not only were these instructors aware of problems with their instruction, but they were also aware of research-based instructional innovations that might be useful in solving the problems. Four of the five instructors were reasonably familiar with Physics Education Research (PER).

They were aware of the names and basic practices involved with innovative curricula as well as a number of more general strategies (Table 2). The fifth instructor (Gary), while not explicitly familiar with PER had been exposed to general research-based teaching techniques through a residential grant-sponsored program. Thus, these instructors appear to have a reasonable degree of knowledge about possible solutions to the instructional problems that they face.

*Making Instructional Changes*

During the interview we asked the instructors to describe instructional changes that they had made and how their knowledge of educational research had influenced these changes, if at all. We were then able to classify each of these self-reported changes on the adoption-invention continuum (Table 3). The classifications were based on the definitions of each category described earlier (Figures 1 and 2). Below is an example for each category to clarify the categorization.

Adoption: Use of CSEM as an assessment instrument

After talking about some of the instructional changes he had made, Harry was asked how he knows whether the changes are working or not. After discussing how he uses informal cues during class time to assess student understanding he commented that he made use of pre-post testing using the Conceptual Survey of Electricity and Magnetism[40] (CSEM): "*CSEM scores are another good thing. If your CSEM scores are good then, you must be, well, of course you could just be teaching to the test, but assuming that you aren't teaching to the test, then presumably you are doing something right.*" (Harry, 125-127) This use of the CSEM comes directly from and is consistent with the use recommended by educational researchers.

Adaptation: Use of Physlets

Barry discussed learning about Physlets after attending a colloquium by one of the developers and coming across them on internet searches and on Merlot.[41] He discussed integrating Physlets into his courses as a source of animations during class (Barry 606-607). Because the Physlet developers present Physlets as a flexible technological resource, pure adoption is not possible. The developers say that Physlets can be used as classroom demonstrations, but leave the instructor to adapt the resource to their own pedagogical strategies.[42] The use of Physlets as a demonstration appears to be based on the educational principle that an animation can help students visualize physics concepts better than static illustrations. Barry indicated that he based his use of Physlets on this principle and did not indicate modifying the available Physlets in any way.

<u>Reconstruction: Use of Small Group Work Involving White Boards to Increase Interactivity</u>

Many of the instructional recommendations based on physics education research focus on ways to make the classroom more interactive and students more mentally active.[43] One suggestion for how to do this involves having students work in groups on small white boards.[44] Mary was aware of the importance of interactivity in the classroom and became aware of the use of white boards through PER. In using white boards, however, she appears to have developed most of the principles and practices herself based on her interest in promoting interactivity, but also her concern with promoting problem solving skills and with motivating students. She described assigning a problem during class for students to solve in their assigned groups on white boards. After students worked for a while "*I take a white board and choose three or four representative ones, bring them to the front of the room and we talk about them as a whole group. What were they trying to do? Why isn't this a valid approach? What is wrong with the picture that made the whole problem not follow correctly from it? This group, where did they get caught? Why is this right? And that seems to help a lot with their motivation because they like seeing their answers up there.*" (Mary, 633-638)

<u>Invention: Soliciting Questions to Encourage Students to Read the Text</u>

Harry was aware that students do not often read the text before coming to class and also that it is important for the instructor to know what difficulties students are having. Although he was aware that Mazur recommended giving pre class reading quizzes to solve this problem,[19] Harry developed his own technique where he asked students to submit a question to him about the reading via email. (Harry 31-33) This is consistent with best practices since it encourages students to think about their own understanding of the topic and provides the instructor with an understanding of how the students are thinking about the topic and specific difficulties that they are encountering.

*Instructors Engage in Informed Invention and Invention*

Notice from Table 3 that most (70%) of the 20 identifiable changes reported by the instructors fell on the reinvention/invention side of the continuum. Half of the changes fell in the reinvention category. This means that the instructors generally agreed with education research on what the problems were and the general idea of the solutions, but did not take the complete research-based solutions and implement them. They developed or substantially changed the principles and details of the solution. The only adaptations were in the use of Physlets and small group work. Both Terry and Barry reported using Physlets as demonstrations in class to help students visualize physics phenomena. Barry also reported having small groups work on homework-like problems based on a similar practice by one of his colleagues. The only adoptions were the three instructors who report using the Force Concept Inventory[39] (FCI) and/or Conceptual Survey of Electricity and Magnetism[40] (CSEM) as assessments in their courses. Although these are not instructional strategies per se, they are important PER products. All instructors reported using these products as recommended. For example, none described using subsets of the tests to evaluate their instruction.

As would be expected, if faculty develop the details of implementation on their own, some are likely to do so in a way that is consistent with the current research-based understanding of teaching/learning and others are likely to do so in a way that is inconsistent. The examples given previously of reinvention and invention were ones that we judged to be consistent with current research-based understanding of teaching and learning. Below, are examples of reinvention and invention that we judged to be inconsistent with the current research base.

Reinvention incompatible with PER: Fundamental Modifications to Peer Instruction

Terry was aware of Peer Instruction and had concerns about his instruction that Peer Instruction was designed to improve (e.g., students don't get much from a traditional lecture). He described using Mazur's and his own ConcepTests in instruction, but using them in a very different way than Mazur advocates. Mazur argues that student-student discussion of ConcepTests is an essential component of Peer Instruction.[19] Terry, however, reports rarely having students discuss with one another. Also, suggesting incompatible use of Peer Instruction, Terry reports that it is quite common for all students to answer a ConcepTest correctly. Mazur, however, suggests that ConcepTests are most effective when the "initial percentage of correct answers is around 50%" (ref [19], p. 12).  This suggests that the questions Terry uses are inappropriately easy (by Mazur's definition) or that the procedure used to determine student understanding is not accurate (he did not describe how he measured student responses to ConcepTests during the interview).

<u>Invention incompatible with PER: Assigning Reading Exercises to Encourage Students to Read the Text</u>

Similar to Harry, Gary was aware that students do not often read the text before coming to class. As a solution to this problem Gary developed a set of 20 to 30 short reading exercises for each assigned chapter that students were expected to complete before the topic was discussed in class. He describes each question as being "*real short, the answers should be right there, and I make up the exercises while I'm looking at the book, to make sure the answer is really available. . . It will be things like ah, maybe a question on the definition of a new concept.  What is specific heat?  or What's the equation that describes specific heat and what's the symbol that's used for specific heat, what are the units for specific heat.  It's just really making them consciously think about that at least once for 15 seconds.  So, yeah, they ought to be able to just breeze right through it.*"  (Gary 51-63)  We describe this as incompatible with PER because the reading exercises do not appear to encourage students to deeply engage with the material or their understanding of it, but rather to focus on surface level details, something that educational research warns against.[45]

**B. Interactions between Educational Researchers and Instructors**

Although all of the instructors reported making some instructional changes and these changes were often precipitated by a general idea from research (i.e., the reinvention mode), most of the research-based resources and knowledge were not used.  Why would instructors engage in reinvention and invention when there is so much good research-based work readily available?  During the interviews it became apparent that these instructors had problems not only with some of the results of education research, but also with the way in which research practitioners disseminated these results.  Many of these instructors expressed great frustration with this situation.  They appear to want different things from the research community than they perceive the community as currently offering.  In the following paragraphs we describe four categories that emerged related to the interactions between researchers and the instructors.

Three of the instructors (Mary, Terry, and Harry) were reasonably familiar with modern PER research. They all articulated their belief that PER expected adoption/adaptation and they indicated a specific resistance to this mode. Thus, many of the barriers discussed here come from their interviews. Barry was aware of some of the PER results and curricula and indicated that he often discussed teaching and learning issues with a PER researcher and colleague who he described as a close personal friend. He did not describe other interactions with the PER community or any knowledge of PER outside of this personal relationship. Gary did not indicate an explicit awareness of any modern PER research or products; however, he appeared to be aware of basic practices of some PER curricula, such as Peer Instruction.

*Category I. PER is perceived as dogmatic*

The interviewed faculty tended to see educational researchers as not really interested in them or their students, but rather as promoting a particular curriculum. Instructors described what they saw as this sales or evangelist mentality of PER practitioners as making their interactions somewhat confrontational. *"The interactions between the two tend to be that the teacher is critical, the education researcher is trying to make a point, and a lot of time the conversation between the two, as soon as it hits a snag . . ., they [the educational researcher] hide behind what feels like a smoke screen."* (Mary 742-746) Instructors also criticized researchers as promoting their instructional package or technique with the expectation that it will work well in any environment, even ones quite different from the one in which it was developed. *"All of those people seem to think that their way is the only way. . . That the only way that a student's going to learn is if I stop doing this and start doing that. And I argue that in fact that's unfair to both teachers and students. I think that in fact that we need to be telling teachers and students is that students learn in many different ways, that teachers teach well in many different ways and that they ought to be trying to find in the things that are presented by some good solid research or whatever ways in which they need to be addressing the students they have in the place where they have them. And I think the one size fits all is not very good for the whole physics community."* (Terry 733-739) Finally, several instructors thought that educational researchers were not being honest with them and that many PER curricula had flaws that were not usually mentioned or don't work as promised.

Harry: I thought it [Peer Instruction] was taking up a large amount of time. I mean I'd be spending most of a class on a couple of these questions.

Interviewer: And, do you know why that is, because that seems different from what Eric Mazur talks about? He talks about taking 2-3 minutes.

Harry: No. Others have tried to apply that here and they'll tell you the same thing. It winds up taking an inordinate amount of time. Now we might not be applying it right. But, you know, I just xeroxed his problems and put them on the overhead. (Harry 532-538)

Some instructors also noted that, even though researchers often present PER as if there is only one way to teach, there are often times when there does not appear to be agreement among researchers as to what constitutes best practices: "*I haven't gone to a completely, just group work, which seems, some of the research seems to indicated that that's the best thing to do. But then again, I've looked at other research and they did only one interactive session per week and seemed to have the same results as other groups who did totally interactive.*" (Mary, 545-549)

*Category II: Perception That PER Says I'm a Bad Teacher*

The research community has put a great deal of effort into discrediting traditional transmissionist instructional approaches. It is not unusual for researchers to report studies where research-based innovations are compared to more traditional lecture-based approaches with the innovation being shown to be superior. While it is likely necessary for faculty to become dissatisfied with the traditional approach to teaching before they will consider alternatives, evidence from this study (see Table 1) and other studies[1] suggest that this dissatisfaction already exists for many faculty. While it is clearly important to emphasize differences between research-based practices and traditional instruction, it appears this approach can be problematic if done without consideration for the emotional reactions that can be engendered.

The faculty we interviewed described emotional reactions to the message of educational researchers. These instructors saw educational researchers as insinuating that they are bad teachers. *"The first word out of their [a typical PER presenter] mouth is you're not doing things right."* (Terry 831-832) *"If you tell me that you think my teaching is bad that automatically sets up a barrier. If I tell you that the only really good way to teach introductory physics is X, I've again set up some kind of barrier. . . I think there's just too much of that going on right now."*

(Terry 954-957) This was often due to researchers who contrast pedagogical practices that they believe faculty commonly engage in with instructional practices that the researcher is promoting. *"Basically what [specific educational researcher] does is . . . gives you a lot of practical help on ways you can go wrong. Don't think this will work because these students are interpreting it this way. I've taken a lot of those things to heart, but I'm not sure I know where to go. You know, it can be paralyzing some of these dicta."* (Harry 440-444)

These faculty care about their students and have done their best with the knowledge they possessed and under the circumstances they found themselves. An important part of their identity is their role as an expert teacher. It is difficult for them when they perceive that the research community is telling them that they've been doing it all wrong and perhaps even causing harm to their students. Not unexpectedly, their reaction can be defensive.

Instead of making them feel they are bad teachers and that they are being told to adopt research innovations because the researchers know best, these instructors would like the research community to recognize that they have valuable experiences and expertise and work with them to improve teaching and learning. *"[I want the research community to say] not that you're doing a bad job, but here are some new ways that this community has discovered about how students learn, ways in which students can learn better, topics in which students have the most serious trouble."* (Terry, 855-857)

These faculty are, in fact, correct. They are expert teachers with a career of experiences who are capable of using their knowledge to integrate research based ideas into their own classrooms. They agree that they can improve their instruction but want their expertise and experiences to be respected. The instructors want to work with the research community to make improvements and not be made to feel their ideas are being judged or discounted.

*Category III: Educational Research Results and Methods are Questioned*

In addition to not necessarily trusting the motives of educational researchers and not feeling that their professional knowledge was validated, faculty also identified many flaws they perceived in educational research methods that they used to justify discounting some results.

<u>Example 1:  PER places too much emphasis on conceptual inventories like the FCI</u>

Instructors criticized the widespread use of conceptual inventories like the FCI and CSEM.  They tended to think that the limited content covered was a significant drawback.  *"The FCI is a very short, very focused topical test made up by a certain group of faculty members who have obviously a certain bias in the process and its got 30 questions on basically Newton's three laws."* (Terry 860-862)  Also, most suggested that the scores were not terribly meaningful since it is very easy to teach to the test.  *"The real flaw, as a scientist, that I see with the FCI is you know what the questions are ahead of time.  So, the potential for teaching to the test is huge.  And, and, consciously or unconsciously being aware of exactly what's going to be asked, just has too big of an influence about how a class is taught.  And it's just too easy, even if a teacher doesn't want to, to overemphasize some detail."* (Mary 493-498)

<u>Example 2: PER uses inappropriate comparison courses</u>

Many instructors criticized the comparison courses used in PER studies.

"All of these studies tend to be done with people who are very concerned about teaching, very interested in it, and are putting a lot of effort into their teaching at the time because they are making changes.  And, all of those things alone could make a really big difference in how effective a class is.  And, so to compare somebody who's doing that to somebody who's doing the usual same old same old with notes that they came up with ten years ago is not really valid.  It's not really fair." (Mary 662-667)

<u>Example 3: PER studies are typically short term</u>

Many instructors appeared to view their job as teachers as one to prepare students to be successful after their course.  "My students I hope are well taught and the only proof of that is not in any test they take or anything else, its where they go and what they do afterwards.  That's my bottom line." (Terry 259-260)  They tended to criticize educational research for its lack of focus on the long term.  "I mean a big problem with a lot of the physics research I have is they never really track to see how it continues and so students who've had an interactive style teaching the beginning, how do they do as  seniors, how do they do on these standardized tests?  How do they do as graduate students?  How does it trickle on up?" (Mary 658-661)

While the critiques these faculty raised have some validity, they can all be reasonably countered. For example, while some PER studies have been done comparing results of courses taught by PER researchers to courses taught by extreme traditionalists, there are many studies which compare innovations across comparable instructors and uphold previous findings. The objections raised by our interviewees, while often true in a narrow frame generally fail to account for the ways in which the findings of PER have been replicated over time.

Do faculty just need to be better informed about the nature of educational research and the ways in which the educational research community has in fact addressed many of their concerns? We cannot fully answer this question from our data, but given the level of knowledge demonstrated by our interviewees about educational research we suspect that simple ignorance cannot completely explain the critiques. It is impossible to conduct a research study to which no critique could be made. It will always be possible that some alternative theory has not been fully discounted or that not all variables have been controlled for. Ultimately it is up to the consumer of the research to decide whether or not to use the results. If the consumer, for whatever reason, does not want to accept the results, he or she will be able to find a critique to justify this decision.

As is often sarcastically noted by educational researchers, science faculty who are well versed in scientific thinking appear to value their intuitive thinking over scientific evidence when it comes to issues of teaching (see, for example Ref 9). The theory of Cognitive Dissonance[46] may be helpful in explaining this seemingly non-scientific rejection of many aspects of educational research. According to the theory of cognitive dissonance and many empirical experiments based on this theory, humans can go to great lengths to avoid dissonance between their behavior and self-concept.[47] Thus, for the instructors in this study to accept educational research in the way it is presented, they would have to also accept the idea that their prior teaching had significant deficits. This threatens their self image as good teachers and is likely to be a particular issue for senior instructors who have spent most of their careers working under a traditional instructional paradigm. One way out of the dissonance is to discredit the research. These thought processes, of course, happen at an unconscious level.

Our speculation that the theory of cognitive dissonance is applicable here is supported by instructor statements discussed above in the section on "PER says I'm a bad teacher." While improvements can certainly be made in research methodology used by researchers and in the communication of this methodology to faculty, from a cognitive dissonance perspective this improved rigor will be unlikely to be significantly more convincing. Faculty are human and react to the findings of research both intellectually and emotionally. The research community has largely failed to acknowledge and address this emotional aspect to the reception of our work.

*Category IV: Faculty Want to Be Part of the Solution*

As a result of the way that these faculty perceive their interactions with educational researchers and the research results themselves, they tend to not make full use of research-based findings. They recognize that research has some good things to offer them and that researchers have expertise in teaching and learning that could be valuable. Yet, they feel a need to be part of the solution themselves. "*I've spent my life doing this [teaching] and part of my teaching is in fact to be aware of all of the things that are going on [in educational research], but I want it to be useful and meaningful to that discourse.*" (Terry 914-916). This results in a situation where the primary way that these instructors pick the good from the bad is to use their own intuition and experience. "*I mean how much time I'm willing to devote to a technique that they've claimed they've shown does something, If I don't feel that the method was scientific enough and that they've demonstrated that it was scientific, they haven't given enough detail, I use much more my gut instinct. Yes, this is something I think would work or No, I don't think this is something that would work.*" (Mary 677-681)

All five instructors described the instructors' personal style, preferences, and skills as being very important in determining appropriate teaching practices. Thus, they did not expect any instructional package created elsewhere to work well for them with minimal or no modifications. This explains why, as described earlier, they did not follow the adoption model even in cases when they believed in the usefulness of the innovation. What most of the instructors seem to describe as a desirable situation is some degree of reinvention where a change agent will work with them to decide on instructional practices that fit their individual situations. This would be based on the instructors' knowledge, skills, preferences, and teaching situation as well as on the available research knowledge about teaching and learning. "*I think what we ought to be doing is we ought to be talking to the teachers in the physics community about all of the possibilities, all of the ways in which students learn*" (Terry 720-724) "*The blanket statement doesn't hit me that some things are better than others because I think what you have to do in that statement is define* *which things are good and which things are bad and which teachers they are good and bad for and which students they are good and bad for.*" (Terry 762-764)  One instructor, Harry, wanted to go even further.  He described his ideal situation as one of invention where PER would provide a coherent conceptual model of teaching and learning so faculty can make their own instructional decisions.  *"There is a problem with physics education research ... I don't have a mental model of how students learn.  . . .If you claim that a certain optical phenomena occurs I can go to my office and calculate that and say, oh, yeah that can happen or no there is no possible way.  I have a good feel for the conditions under which that occurs.  I can't do the same thing for students, for their learning.  I don't have an intellectual framework around which to organize  innovations in teaching . . .So all I can do is to try and implement what I see in the literature as best practices."*  (Harry 95-101)  *"So, how could PER be of more use to m, if you could come up with answers to questions like that.  What methods of presentation, be they textual or electronic, or whiz-bang, or whatever, what methods of presentation most effectuate learning.  And what kinds of learning.  If I had a framework like that then I could answer my own questions. . . . OK, I want to do this.  Here's how the experts tell me, here's the things that the experts tell me I have to consider, OK I'll consider it.*" (Harry 612-620)

## VI. Discussion

These faculty are aware of research-based products and generally agree with educational researchers about the problems those products are designed to solve.  There does not appear to be a difficulty with awareness.  The research community appears to have been effective in this level of communication.  The purposeful nature of this sample, however, should create caution in generalizing this result.  Several of the interviewees indicated that they had always been aware of the problems that the research-based curricula were designed to solve.  Thus, they may have been predisposed to pay attention to the research findings.  This would be consistent with the results of a study of the dissemination of educational research to mathematics faculty[33] which concluded that "the only interviewees [math instructors] who were open to being persuaded by the dissemination materials were those who were already interested in or committed to the need for a reform but who were shopping around for the right approach." (Ref 33, p.11)  Nonetheless, even though the instructors we interviewed may have started to pay attention to research findings because they were interested in reform, the research community appears to have sustained this interest.

These faculty view educational researchers as expecting the change agent-instructor interaction to follow the adoption model. Yet, they think that the most productive change agent-instructor interaction would be for researchers and faculty to work together under the reinvention or invention model. In practice, though, faculty tended to work alone under the reinvention model. In some cases faculty reinvented instruction that was consistent with the original intention and/or recommendations in the research literature, but in many cases they reinvented instruction that was missing important fundamental features of the intended instruction and/or conflicted with recommended practices. This isolation also meant that there was no sharing of successes or failures so that others could learn from them.

Although this study was limited to a purposeful sample of five college physics instructors, many of the same themes were found by a different group of researchers in a study of a college biology instructor's interactions with an instructional reform program.[48] Quotes and analysis in the article exhibit some of the same emotional reactions to the instructor's perception that change agents are telling her that there is only one good way to teach, that she is a bad teacher if she does not teach that way, and that her professional knowledge and experience are not valued. Ultimately, this instructor behaved similarly to the physics instructors described in this paper. She took some aspects of the reform ideas that she thought were useful and incorporated them into her preexisting instructional style while rejecting others (i.e., the reinvention mode). She did not, though, make fundamental changes to an inquiry mode of instruction which was a primary goal of the reform program. The presence of this phenomenon in a different context (instructional reform in college-level biology) suggests that the issue of divergent expectations between change agents and college science faculty may be broadly applicable. Further research is needed to determine the extent of divergent expectations as a barrier to instructional change in the general population of science faculty.

## VII. Implications and Recommendations

These faculty were a purposeful sample of the most likely users of educational research. Although there are individual cases where faculty have adopted a research product more or less as is, it is likely that the reinvention and invention tracks identified here are much more common.[49] Faculty tended to work alone in their reinventions and inventions even though they would have liked to work more closely with the research community. They all indicated that adoption wouldn't work because of the personal nature of teaching and the unique instructional environments. Many instructors also offered a more emotional reason – that they felt researchers implied they were a bad teacher and did not recognize the value of their work and experience. This issue of 'face' has come up in the research literature on dissemination of innovations.[50]

It appears that the educational research community may have a broader impact on actual teaching practices by more fully embracing a mode of interaction with traditional faculty based on cooperation, respect, and support. Instructors are not simply "teaching technicians," they want to, and should be included as active participants in the development process. Before we can help faculty to reinvent/invent, we must first gain a better understanding of the conditions under which this can be done successfully and make this understanding an explicit part of the dissemination process. In this section we offer some speculative recommendations about how the research community might begin to move in this direction.

## Recommendations

*Recommendation 1: Provide easily modifiable materials.*

Moving towards the invention side of the adoption-invention continuum means that instructional materials and designs should be developed with the expectation that faculty will engage in local customization. Faculty should be treated as participants in the development process and should be given the opportunity to adopt materials for their local environment. In addition, providing instructors with easily modifiable materials communicates to them that they can and should use their own expertise to appropriately integrate the materials into their unique teaching situations. One example is a project recently undertaken by Andy Elby and the University of Maryland Physics Education Research Group that encourages and supports customization through easily edited materials along with explanations about the instructional design and annotated video snippets of the materials and techniques in actual classroom use.[51]

*Recommendation 2: Disseminate and research ideas in addition to curriculum.*

If faculty are going to modify curriculum effectively, they need to understand both *what* works (details) as well as *why* it works (principles). For example, while many physics faculty now have a copy of Mazur's Peer Instruction[19] and may have begun using some of the associated conceptual questions, they are less aware of the research evidence that learning is primarily a social activity[52] and, so, tend to drop the peer-peer interaction part of Peer Instruction.[54] Without an understanding of the social importance of learning, it is then easy for an instructor to reinvent peer instruction in a way that is likely to reduce its effectiveness. On the other hand, once an instructor does understand the importance of social interactions for learning, they are more likely to incorporate this aspect into their own reinventions or inventions.

In order to provide faculty with the details and principles knowledge, the educational research community will need to better understand and clearly articulate why a curriculum is successful and not just document its success at one, or a handful, or institutions.

*Recommendation 3: Explicitly research the conditions for transfer.*

It is not uncommon for curriculum to be produced and disseminated that has not been tested in contexts beyond the environment in which it was developed. Most research-based curricula has been developed at research universities or elite liberal arts colleges. However, both conventional wisdom and available evidence[27,55,56] suggest that these curricula do not always transfer directly to other environments. In order for dissemination to be successful we suggest that curriculum development efforts: (1) test and refine curriculum in environments fundamentally different from the development site; (2) attempt to make explicit what aspects of the curriculum will transfer and under what conditions the transfer will be successful; (3) make recommendations for modifications in different contexts, for example, how the curriculum could be modified for different sized classes, or for schools with less prepared students; (4) articulate why some aspects transfer better than others to guide instructors in their modifications. Understanding the whys behind transfer issues is also essential for building a general model to guide future development projects.

*Recommendation 4: View faculty as partners.*

When disseminating educational innovations, the research community should focus on working with faculty as partners, either individually or in small groups to improve instructional practices in individual situations. Under this framework faculty would be recognized as a valuable part of this process with learning occurring on both sides. This is in contrast to current dissemination activities describing deficiencies with traditional instructional practices, providing polished ready-to-use curricula, and having change agents promote only the curricula that they developed.

While constructivist principles are well recognized and supported for students learning physics, these same ideals are often replaced by a "teaching by telling" approach to dissemination. Not only are the general beliefs, previous experiences, and intuitions of instructors important influences in the way they integrate new approaches, their expertise is also valuable and should be recognized as such. Such an approach would avoid the issues of face and cognitive dissonance because it would use the instructor's current instruction as a starting point for constructing new instruction as opposed to requiring the instructor to first reject his/her current instruction. There are many ways to do this, ranging from one-on-one interaction[57-59] to more formally organized groups of faculty interested in improving their instruction.[60-62]

*Recommendation 5: Acknowledge that change is difficult and support, rather than blame instructors.*

Many of the reforms suggested by educational research are difficult to implement. Yet, many innovations are presented as if significant improvements are possible by following the "simple" suggestions of the curriculum developer. In reality, it is common for instructors to try and integrate a new research-based idea into their teaching without a noticeable improvement in instruction. Often, the reforms fail in large part because of situational constraints. For example, most research-based curricula require students to interact with each other. However, giving students the opportunity to speak in class slows down the pace creating a difficulty for instructors without the ability to reduce the content that is covered. Also, if students do not talk to each other in any other courses they may be particularly resistant, creating an atmosphere that resists the innovation.

Educational structures have developed around and in support of traditional instructional practices.[14] Reformed instruction is necessarily instruction that in some way challenges the status-quo. The greater the challenge to the status-quo the innovation represents, the greater the resistance that can be expected. Too often, an instructor may try an innovation and then blame the method for the poor results. In return, it is common for the research community to blame the instructor for the failure. Rather than viewing failed implementations as entirely the fault of the implementers, it would be useful for the research community to acknowledge how difficult real and sustained change can be, and to identify and articulate the situational factors that make such change difficult. In addition, instructors should be made aware of these difficulties so they can better work to overcome them. Finally, the research community should provide supportive structures to help faculty to cope with the barriers they are likely to encounter as they try to make improvements in their instruction.

## VIII. Acknowledgements

The authors would like to thank the five instructors who gave their valuable time to participate in this study as well as Leon Hsu, Robert Poel, David Rudge, and David Schuster for their helpful comments on earlier versions of this manuscript.

## IX. References

**1**    Yerushalmi, E., Henderson, C., Heller, K., Heller, P. and Kuo, V. (2006) Physics faculty beliefs and values about the teaching and learning of problem solving part I: Mapping the common core. *To appear in Physical Review Special Topics: Physics Education Research*

**2**    Prosser, M. and Trigwell, K. (1999) Understanding learning and teaching: The experience in higher education, St. Edmundsbury Press.

**3**    Peer Instruction is perhaps the most widely used research based instructional strategy. A widely disseminated survey was able to identify 384 self-described users of Peer Instruction or similar strategies [see Ref 4]. Respondents were from a "broad array of institution types" throughout the world and "most" used Peer Instruction to teach physics. However, even if we assume that all 384 users of Peer Instruction teach tertiary-level physics they would represent just 3.4 % of the roughly 11,360 physics faculty employed in two-year and four-year colleges in the United States. Physics faculty data can be found in references 5 and 6.

**4**    Fagen, A. P., Crouch, C. H. and Mazur, E. (2002) Peer instruction: Results from a range of classrooms. *The Physics Teacher* 40, 206-209.

**5**    Ivie, R., Stowe, K. and Nies, K. (2003) *2002 physics academic workforce report* (AIP Pub. Number R-392.5) American Institute of Physics.

**6**    McFarling, M. and Neuschatz, M. (2003) *Physics in the two-year colleges: 2001-02* (AIP Pub. Number R-436) American Institute of Physics.

**7**      National Science Foundation. (1996) Shaping the future: New expectations for undergraduate education in science, mathematics, engineering, and technology.  A report on its review of undergraduate education by the advisory committee to the national science foundation directorate for education and human resources. (NSF 96-139) National Science Foundation.

**8**      National Research Council. (2003) Improving undergraduate instruction in science, technology, engineering, and mathematics: Report of a workshop, The National Academies Press.

**9**      Handelsman, J., Ebert-May, D., Beichner, R., Bruns, P., Chang, A., DeHaan, R., Gentile, J., Lauffer, S., Stewart, J., Tilghman, S. M. and Wood, W. B. (2004) Education: Scientific teaching. *Science* 304 (5670), 521-522. (Available online: http://www.sciencemag.org)

**10**     Henderson, C. and Stelzer, T. (2005) The gap between PER and mainstream faculty: The PER perspective. Poster presented at the Foundations and Frontiers in Physics Education Research Conference, Bar Harbor, Maine, August 16, 2005. (Available online: http://homepages.wmich.edu/~chenders/Publications/FFPER05Poster.pdf)(

**11**     Henderson, C., Stelzer, T., Hsu, L. and Meredith, D. (2005) Maximizing the benefits of physics education research: Building productive relationships and promoting institutional change. *American Physical Society Forum on Education Newsletter* (Fall 2005), 11-14. (Available online: http://homepages.wmich.edu/~chenders/Publications/HendersonFEDFall2005.pdf)

**12**     Redish [Ref 13] identifies 5 features of a traditional physics course in the USA. These have been modified slightly from his original list.  (1) It is content oriented.  (2) It has 3-4 hours of lecture and 0-1 hours of problem solving recitation per week.  (3) If there is a laboratory, it will be 2-3 hours and "cookbook" in nature.  (4) The instructor is active during the class session and students are passive.  (5) The instructor expects the student to undergo active learning activities outside of the class section, in reading, problem solving, etc., but does not usually enforce these activities.

**13**     Redish, E. F. (2003) *Teaching physics with the physics suite*, John Wiley & Sons.

**14**     Dancy, M. and Henderson, C. (2005) Beyond the individual instructor: Systemic constraints in the implementation of research-informed practices. In *Proceedings (peer reviewed)* *of the 2004 AAPT Physics Education Research Conference* (Vol. 790) (Franklin, S. Marx, J. and Heron, P., eds.), American Institute of Physics. (Available online: http://homepages.wmich.edu/~chenders/Publications/PERC2004Dancy.pdf)

**15**      Henderson, C. and Dancy, M. (2005) Teaching, learning and physics education research: Views of mainstream physics professors. In *Proceedings (peer reviewed) of the 2004 AAPT Physics Education Research Conference* (Vol. 790) (Franklin, S. Marx, J. and Heron, P., eds.), American Institute of Physics. (Available online: http://homepages.wmich.edu/~chenders/Publications/PERC2004Henderson.pdf)

**16**      Tannen, D. (1990) You just don't understand: Women and men in conversation, Ballantine Books.

**17**      Tannen, D. (1986) That's not what I meant! How conversational style makes or breaks your relations with others, Ballantine Books.

**18**      Slater, T. F. (2003) When is a good day teaching a bad thing? *The Physics Teacher* 41 (7), 437-438.

**19**      Mazur, E. (1997) *Peer instruction: A user's manual*, Prentice Hall.

**20**      Felder, R. M. and Brent, R. (1996) Navigating the bumpy road to student-centered instruction. *College Teaching* 44, 43-47. (Available online: http://www.ncsu.edu/felder-public/Papers/Resist.html)

**21**      Clement, J. (2000) Analysis of clinical interviews: Foundations and model viability. In *Handbook of research design in mathematics and science education* (Kelly, A.E. and Lesh, R., eds.), Lawrence.

**22**      Ellsworth, J. B. (2000) *Surviving change: A survey of educational change models* (Report No. IR-109) Office of Educational Research and Improvement.

**23**      Fullan, M. (2001) *The new meaning of educational change*, Teachers College Press.

**24**      Rogers, E. M. (1995) *Diffusion of innovations*, Free Press.

**25**      Creswell, J. W. (1998) Qualitative inquiry and research design: Choosing among five traditions. Sage.

**26**    Henderson, C. (2005) The challenges of instructional change under the best of circumstances: A case study of one college physics instructor. *Physics Education Research Section of the American Journal of Physics* 73 (8), 778-786.

**27**    Saul, J. M. and Redish, E. F. (1997) Final evaluation report for FIPSE grant #P116P50026: Evaluation of the workshop physics dissemination project University of Maryland.

**28**    Spillane, J. P. (2004) Standards deviation: How schools misunderstand educational policy, Harvard University Press.

**29**    Stigler, J. W. and Hiebert, J. (1999) The teaching gap: Best ideas from the world's teachers for improving education in the classroom, The Free Press.

**30**    Wallace, C. S. and Kang, N.-H. (2004) An investigation of experienced secondary science teachers' beliefs about inquiry: An examination of competing belief sets. *Journal of Research in Science Teaching* 41 (9), 936-960.

**31**    Wittmann, M. C. (2002) *On the dissemination of a proven curriculum: Realtime physics and interactive lecture demonstrations* (White paper web publication of FIPSE external evaluator report for the RTP/ILD dissemination project)

**32**    Seymour, E. (2001) Tracking the process of change in us undergraduate education in science, mathematics, engineering, and technology. *Science Education* 86, 79-105.

**33**    Foertsch, J., Millar, S. B., Squire, L. and Gunter, R. (1997) *Persuading professors: A study of the dissemination of educational reform in research institutions*, Report to the NSF Education and Human Resources Directorate, Division of Research, Evaluation, and Communication, Washington DC. University of Wisconsin-Madison, LEAD Center.

**34**    Driel, J. H. v., Beijaard, D. and Verloop, N. (2001) Professional development and reform in science education: The role of teachers' practical knowledge. *Journal of Research in Science Teaching* 38 (2), 137-158.

**35**    Winter, D., Lemons, P., Bookman, J. and Hoese, W. (2001) Novice instructors and student-centered instruction: Identifying and addressing obstacles to learning in the college science laboratory. *The Journal of Scholarship of Teaching and Learning* 2 (1), 15-42.

**36**    Hutchinson, J. and Huberman, M. (1993) Knowledge dissemination and use in science and mathematics education: A literature review (NSF 93-75) National Science Foundation.

**37**    National Science Foundation. (2005) Course, curriculum, and laboratory improvement (CCLI): A solicitation of the division of undergraduate education (DUE) (NSF 05-559) National Science Foundation.

**38**    References to interview statements used throughout this paper include a pseudonym to identify the interviewee and the line number(s) of the statement in the interview transcript.

**39**    Hestenes, D., Wells, M. and Swackhamer, G. (1992) Force concept inventory. *The Physics Teacher* 30 (3), 141-151.

**40**    Maloney, D., O'Kuma, T., Hieggelke, C. J. and Van Heuvelen, A. (2001) Surveying students' conceptual knowledge of electricity and magnetism. *American Journal of Physics* 69 (7), S12-S23.

**41**    Multimedia Educational Resource for Learning and Online Teaching (MERLOT) is a free and open resource designed primarily for faculty and students of higher education. Links to online learning materials are collected here along with annotations such as peer reviews and assignments. (URL: http://www.merlot.org/Home.po)

**42**    Christian, W. and Belloni, M. (2004) Physlet physics: Interactive illustrations, explorations, and problems for introductory physics, Pearson Education.

**43**    Hake, R. R. (1998) Interactive-engagement vs. Traditional methods: A six-thousand-student survey of mechanics test data for introductory physics courses. *American Journal of Physics* 66, 64-74.

**44**    MacIsaac, D. (Online Resource: http://physicsed.buffalostate.edu/AZTEC/BP_WB/) *Whiteboarding in the classroom*, The Arizona K-12 Center and the Arizona Teacher Excellence Coalition (AzTEC) Project.

**45**    Chi, M. T. H. and VanLehn, K. A. (1991) The content of physics self-explanations. *The Journal of The Learning Sciences* 1 (1), 69-105. (Available online: http://links.jstor.org/sici?sici=1050-8406%281991%291%3A1%3C69%3ATCOPS%3E2.0.CO%3B2-P)

**46**    Festinger, L. (1957) *A theory of cognitive dissonance*, Stanford University Press.

**47**    Griffin, E. (1997) *A first look at communication theory*, McGraw-Hill.

**48**    Weiss, T. H., Feldman, A., Pedevillano, D. E. and Copobianco, B. (2003) The implications of culture and identity: A professor's engagement with a reform collaborative. *International Journal of Science and Mathematics Education* 1, 333-356.

**49**    Of course, many instructors (perhaps the majority) do not change their instruction at all.

**50**    Zaltman, G. and Duncan, R. (1977) *Strategies for planned change*, John Wiley and Sons.

**51**    McCaskey, T., Hodges, R. and Elby, A. (2005) Integrating adaptable tutorials with professional development materials. Paper presented at the American Association of Physics Teachers Winter Meeting, Albuquerque, NM.

**52**    Based on the ideas of Vygotsky (Ref 53), Redish's fifth instructional principle for physics instruction is that "For most individuals, learning is most effectively carried out via social interactions." (Ref 13, p. 39).

**53**    Vygotsky, L. S. (1978) Mind and society: The development of higher mental processes, Harvard University Press.

**54**    Henderson, C. and Dancy, M. (2005) When one instructor's interactive classroom activity is another's lecture: Communication difficulties between faculty and educational researchers. Paper presented at the American Association of Physics Teachers Winter Meeting, Albuquerque, NM.

**55**    Sabella, M. and Bowen, S. (2003) Physics education research with special populations: How do we characterize and evaluate the special needs and resources of students who are underrepresented in stem education. Poster Presented at the AAPT Physics Education Research Conference (Madison, WI).

**56**    Sabella, M. and Cochran, G. (2004) Evidence of intuitive and formal schemas in student responses: Examples from the context of dynamics. In *Proceedings (peer reviewed) of the 2003 Physics Education Research Conference* (Vol. 720) (Franklin, S. Cummings, K. and Marx, J., eds.), pp. 89-93, American Institute of Physics.

**57** Briscoe, C. and Prayaga, C. S. (2004) Teaching future k-8 teachers the language of newton: A case study of collaboration and change in university physics teaching. *Science Education* 88 (6), 947-969. (Available online: http://dx.doi.org/10.1002/sce.20005)

**58** Van Sickle, M. and Kubinec, W. (2002) Transforming teaching: A physics professor's thoughts. *Journal of College Science Teaching* 32 (4), 258-263.

**59** Henderson, C., Beach, A. and Famiano, M. (2006) Creating lasting reform: Induction by co-teaching. Poster presented at the Conference on the Preparation of Physics and Physical Science Teachers, Fayetteville, AR March 24, 2006.

**60** Cox, M. D. (2004) Introduction to faculty learning communities. In *Building faculty learning communities: New directions for teaching and learning, no. 97* (Vol. 2004) (Cox, M.D. and Richlin, L., eds.), pp. 5-23, Jossey-Bass. (Available online: http://dx.doi.org/10.1002/tl.129)

**61** Campbell, D. K., Elliot, C. M. and Gladding, G. E. (1997) Parallel parking an aircraft carrier: Revising the calculus-based introductory physics sequence at Illinois. *Forum on Education Newsletter of the American Physical Society* (Summer), 9-11. (Available online: http://www.aps.org/units/fed/newsletters/aug97/index.cfm#campbell)

**62** Sunal, D. W., Hodges, J., Sunal, C. S., Whitaker, K. W., Freeman, L. M., Edwards, L., Johnston, R. A. and Odell, M. (2001) Teaching science in higher education: Faculty professional development and barriers to change. *School Science and Mathematics* 101 (5), 246-257.

**Figure Captions**

FIGURE 1: Adoption-Invention Continuum

FIGURE 2: Change Agent (CA) and Instructor (I) roles in developing and implementing new instructional strategies.

FIGURE 3: NSF CCLI model of educational change. [From Ref 37]

TABLE 1: Instructional problems discussed by interviewees during interview.

| Instructional Problem | Terry | Harry | Mary | Gary | Barry |
|---|---|---|---|---|---|
| Students don't get much from traditional lecture. | ♦ | ♦ | ♦ | ♦ | ♦ |
| Different kinds of students learn differently. | ♦ | | | ♦ | ♦ |
| Students have misconceptions that are not simple to change. | ♦ | ♦ | | ♦ | |
| Many students have poor problem solving skills. | | ♦ | ♦ | ♦ | |
| Assessment difficulties – getting the right answer to a problem does not mean that a student understands. | | | ♦ | | ♦ |
| In teaching, it is helpful to tailor explanations to individual students, but this is difficult/impossible in a large class | | ♦ | | | |
| Students have great difficulty learning the basic concepts of physics | | ♦ | | | |

TABLE 2: Research-based instructional methods spontaneously discussed by interviewees.

[•]: Mentioned - instructor mentioned name or defining feature of an instructional strategy.

[♦]: Described - instructor mentioned name or defining feature plus at least one additional substantive aspect of the associated instructional activities.

| Research-Based Instructional Strategy | Terry | Harry | Mary | Gary | Barry |
|---|---|---|---|---|---|
| Peer Instruction | ♦ | ♦ | ♦ | ♦ | ♦ |
| Physlets | ♦ | | ♦ | | ♦ |
| Small group work | | ♦ | ♦ | | ♦ |
| Workshop Physics | ♦ | ♦ | | | • |
| Washington Tutorials | ♦ | ♦ | | | |
| Problem solving framework | | • | • | | |
| Personal response systems | | | • | | • |
| Real-Time Physics and Interactive Lecture Demonstrations | ♦ | | | | |
| "Army" method. Pose question, pause, and call on student. | | | | ♦ | |
| White boards to encourage students to interact during class. | | | ♦ | | |
| Physics by Inquiry | | ♦ | | | |
| Student-Centered Activities for Large Enrollment Undergraduate Programs (SCALE-UP) | | | | | ♦ |
| Modeling and discussing expert thinking related to problem solving. | | | ♦ | | |
| Individual interviews with each student – to have motivational personal contact. | | | • | | |
| Have students write down answer after posing a question. | | | | • | |
| Discussion-based teaching techniques | | | | • | |
| Consortium of Upper-Level Physics Software (CUPS) | • | | | | |
| Comprehensive Unified Physics Learning Environment (CUPLE) | • | | | | |
| Matter and Interactions | | | | | • |

TABLE 3: Instructional innovations.  A: Adoption, D: Adaptation, R: Reinvention, I: Invention. Instructors were classified in the most appropriate category based on the available evidence in the interview transcripts.

| **Instructional Strategy** | **Terry** | **Harry** | **Mary** | **Gary** | **Barry** |
|---|---|---|---|---|---|
| Peer Instruction | R | R | R | R | R |
| FCI/CSEM as an assessment instrument | A | A | A | | |
| Small group work | | R | R | | D |
| Physlets | D | | | | D |
| "Army" method. Pose question, pause, and call on student. | | | | R | |
| Discussion-based teaching techniques | | | | R | |
| Modeling and discussing expert thinking related to problem solving. | | | R | | |
| Different instruction for different student abilities. | | | | I | |
| "Exercises" to guide students through solving a problem. | | | | I | |
| Solicits questions from students | I | | | | |
| Lecture-based questions | | | | | I |

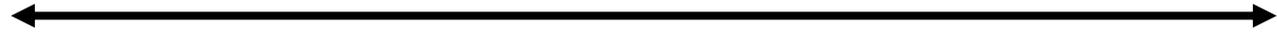

| **Adoption** | **Adaptation** | **Reinvention** | **Invention** |
|---|---|---|---|
| The change agent develops all of the materials and procedures and gives them to the instructor to implement as is. | The change agent develops the materials and procedures and gives them to the instructor who modifies some of the details before implementation. | The instructor uses the ideas or materials of the change agent but changes them significantly (i.e., changes a principle) or develops fundamentally new procedures or materials based on the change agent ideas. | The instructor develops materials and procedures that are fundamentally based on his/her own ideas. |

FIGURE 1: Adoption-Invention Continuum

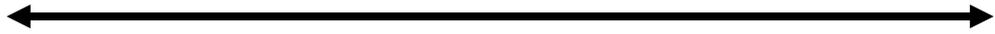

| Activity in the Change Process | Adoption | Adaptation | Reinvention | Invention |
|---|---|---|---|---|
| Identify an instructional problem | CA | Either CA or I | Either CA or I | I |
| Develop general idea of a solution (awareness knowledge) | CA | CA | CA | I |
| Develop principles of the solution (principles knowledge) | CA | CA | I (with or without CA help) | I |
| Develop details of the solution (how-to-knowledge) | CA | I (with or without CA help) | I (with or without CA help) | I |
| Implement solution | I | I | I | I |

FIGURE 2: Change Agent (CA) and Instructor (I) roles in developing and implementing new instructional strategies.

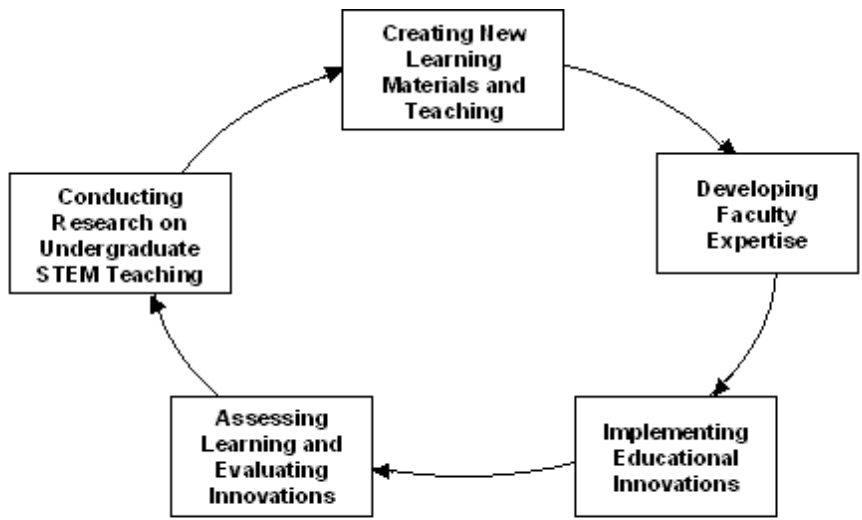

Cyclic model for knowledge production and improvement of practice in undergraduate STEM education

FIGURE 3: NSF CCLI model of educational change. [From Ref. 37]